\documentstyle[aps,prl,multicol]{revtex}
\begin{document}
\draft
\title{
Exact Charge Dynamics in
the Supersymmetric $t$-$J$ Model\\
with Inverse-Square Interaction}
\author{
Mitsuhiro Arikawa,$^1$ Takashi Yamamoto,$^2$ Yasuhiro Saiga,$^3$ and
Yoshio
Kuramoto$^1$
}
\address{$^1$Department of Physics, Tohoku University, Sendai 980-8578,  
Japan\\
$^2$Max-Planck-Institut f\"{u}r Physik komplexer Systeme,N\"{o}thnizer  
Str.
38, D-01187 Dresden,
Germany\\
$^3$Institute for Solid State Physics, University of Tokyo, Roppongi
7-22-1, Tokyo 106-8666, Japan}

\maketitle

\begin{abstract}
Dynamical charge structure factor $N(Q,\omega)$ with $Q$ smaller
than the Fermi wave number
is derived analytically for the one-dimensional supersymmetric $t$-$J$
model with $1/r^2$ interaction.
Strong spin-charge separation in dynamics is found.
$N(Q,\omega)$ with given electron density does not depend on the spin
polarization for arbitrary size of the
system.
In the thermodynamic limit, only two holons and one antiholon contribute
to
$N(Q,\omega)$.
These results together with  generalized ones for the SU($K$,1)
supersymmetric {\it t-J} model
are derived via solution of the Sutherland model in the strong coupling
limit.
\end{abstract}

\pacs{75.10.Jm, 05.30.-d, 71.27.+a}
\begin{multicols}{2}
Strong correlation between one-dimensional electrons has drastic
consequences to their physical properties.
The most spectacular is the spin-charge separation; the elementary
excitations are spinons and holons both of which obey the fractional
statistics \cite{Haldane}.
This is in contrast with the fermionic Landau quasi-particles realized
in higher dimensions.
In the supersymmetric {\it t-J} model with $1/r^2$ interaction
\cite{KY}, these fractional particles appear in the simplest manner.
Exact thermodynamics for the model \cite{KK95J,KK96} can be
interpreted in terms of free spinons and holons.
The charge susceptibility, for example, depends on the electron
density, but is independent of the spin polarization.  This kind of
independence is referred to as the strong spin-charge separation
\cite{KK95J,KK96}.
It is natural to ask how the spin-charge separation appears in
dynamical correlation functions.

In the absence of charge degrees of freedom, the supersymmetric {\it
t-J} model is reduced to the spin chain called the Haldane-Shastry model
\cite{hs1,hs2}. 
For the latter model Haldane and Zirnbauer \cite{Haldane-Zirnbauer} 
derived
exactly the
dynamical structure factor which is non-zero only over a
compact region in the momentum-frequency plane.
Such region is called the support.
For the supersymmetric {\it t-J} model, 
Ha and Haldane \cite{HaHaldane} analyzed numerical results for 
finite-size
systems, and found that only a few number of elementary excitations
contribute to
spectral functions, namely dynamical structure factors.
They proposed a compact support for each spectral function
in the thermodynamic limit, but did not obtain the spectral functions
themselves.
Analytical knowledge should obviously provide deeper insight into strong
correlation effects on dynamics.
Recently, Kato partially derived a spectral function corresponding
to the removal of a particle from the high-density limit \cite{Kato}. 
So far all analytical work in the literature does not include charge
degrees of freedom in the ground state,
and derivation of charge dynamics has remained a theoretical challenge.

In this Letter, we report on exact analytical results on charge dynamics 
for the
one-dimensional supersymmetric {\it t-J} model with $1/r^2$
interaction at zero temperature, and its generalization from SU(2,1) to
SU($K$,1)
supersymmetry.
We show explicitly how the strong spin-charge separation manifests in
charge dynamics.
The charge correlation function is relevant to dielectric and elastic
responses of real systems with quasi-one-dimensional electrons.
Our work should  motivate further analytical work toward other dynamical
correlation functions which are relevant to various spectroscopies
such as neutron scattering and photoemission.

We consider the supersymmetric $t$-$J$ model given by
\begin{eqnarray}
{\cal H}_{tJ}
&=&
{\cal P}
\sum_{i<j}
\left[-t_{ij}\sum_{\sigma=\uparrow,\downarrow}
\left( c^{\dagger}_{i\sigma}c_{j\sigma} + h.c. \right)
\right.
\nonumber \\
& &
\left.
+J_{ij}
\left(\mbox{\boldmath $S$}_{i}\cdot\mbox{\boldmath $S$}_{j}
-\frac{1}{4} n_{i}n_{j}\right)
\right]
{\cal P},
\label{tj-hamiltonian}
\end{eqnarray}
where $c_{i\sigma}$ is the annihilation operator of an electron at
site $i$ with spin
$\sigma$, $n_{i}$ is the number operator and $\mbox{\boldmath
$S$}_{i}$ is the spin
operator.
The projection operator
${\cal P}$
excludes double occupation at each site.
The transfer energy $t_{ij}$ and exchange one $J_{ij}$ are given by
$t_{ij}=J_{ij}/2=t D^{-2}_{ij}$
where $D_{ij} = (N/{\pi})\sin \left( \pi\left(i-j\right)/N\right)$
with $N$ being the number of lattice sites.
This model is generalized to the SU($K$,1) supersymmetry as
\begin{equation}
{\cal H}_{tJ}^{(K)}
= t \sum_{i<j} D_{i j}^{-2} \tilde{P}_{ij} .
\label{K-tj-hamiltonian}
\end{equation}
Here we have introduced a graded permutation operator
$\tilde{P}_{ij} = \sum_{\alpha,\beta} X_i^{\alpha \beta} X_j^{\beta
\alpha}
\theta_{\beta},
$
where $X_j^{\beta \alpha}$ changes a state  $\alpha$ to $\beta$ at
site $j$,
with $\alpha,\beta$ being either $0$ (hole state), or one of $\sigma$ 
=
$1$,$\cdots$,$K$ (spin state).
The sign factor $\theta_{\beta}$ is $-1$ if $\beta = 0$ and $1$
otherwise.
In the case of $K=2$ the expression (\ref{K-tj-hamiltonian}) is
reduced to eq.(\ref{tj-hamiltonian}).

In terms of the density fluctuation operator
$n_Q = N^{-1/2} \sum_{\sigma=1}^K \sum_{l=1}^N X_l^{\sigma \sigma} 
\exp
(iQl)$ with $Q>0$,
the dynamical charge structure factor is defined by
\begin{equation}
N \left(Q,\omega \right) =\sum_{\nu} | \langle \nu | n_Q | 0 \rangle 
|^2
\delta \left( \omega -E_{\nu}+E_0 \right),
\end{equation}
where
$ | \nu \rangle$ denotes a normalized eigenstate of the Hamiltonian
(\ref{K-tj-hamiltonian}) with
energy $E_{\nu}$.

In the SU($K$,1) {\it t-J} model, a holon has the spectrum
$\epsilon_{\rm h}(q) = K t q [2\pi (1-\bar{n})/K+q]/2$ with $\bar{n}$ 
being
the average electron number per site, and another charge excitation 
called
the antiholon has the spectrum
$\epsilon_{\rm a}(q) =  t q [ 2 \pi (1-\bar{n}) -q]/2$
\cite{KK96,HaHaldane}.
The latter represents the excitation from the condensate of holons.
We first give our main result of this Letter, and then its derivation.
The system has $N_{\sigma}$ electrons with spin $\sigma$.
In the small momentum region $0 < Q \le k_F^{\rm Min}$, with
$k_{F}^{\rm Min}$ being the smallest Fermi number of electrons, {\it 
i.e.},
$k_{F}^{\rm Min} = \pi {\rm Min} (N_1,\cdots N_K)/N $,  we derive
$N(Q,\omega)$ in the form:
\begin{eqnarray}
\lefteqn{
N(Q,\omega)
=
A_0  Q^2  \int_0^{2\pi (1-\bar{n})}d q' \
\prod_{i=1}^K \ \int_0^{2\pi/K} d q_i
}
\nonumber \\
& &
\times
\delta
\left( Q - q' - \sum_{i=1}^K q_i\right) \
\delta
\left( \omega - \epsilon_{\rm a} (q') -\sum_{i=1}^K
\epsilon_{\rm h} (q_i)
\right)
\nonumber \\
& &
\times
\frac{\prod_{i<j} | q_i -q_j |^{2/K}}{ \prod_{i=1}^K(Kq_i +q' )^2} \
\epsilon_{\rm a} (q') ^{K-1} \
\prod_{i=1}^K  \epsilon_{\rm h} (q_i)^{1/K-1},
\label{kfn}
\end{eqnarray}
where $A_0 = {K}/({2 \pi})  \prod_{j=1}^K \Gamma(1/K)/ \Gamma
(j/K)^2$.
Surprisingly this $N(Q,\omega)$ has almost the same form as that of
the spinless Sutherland model with the coupling parameter $1/K$
\cite{Ha,Lesage}.
In the latter model excitations consist of $K$ particles created above
the pseudo Fermi level with charge $-e$, and a hole with charge
$K e$.
The only difference in $N(Q,\omega)$ between the two models is the
integration range of momenta.
Thus the result above can be interpreted in terms of the fractional
statistics \cite{KYA}.

With $K=2$, it can be shown that $N(Q,\omega)$ diverges as
$[\epsilon_{\rm h}(Q)-\omega]^{-1/2}$
as the frequency approaches the upper edge corresponding to the holon
dispersion\cite{Mucciolo}.
The static structure factor $N(Q)=\int d \omega N(Q,\omega) $
reproduces the known
result\cite{Gebhard} obtained for the Gutzwiller wave function.
This should be so since the exact ground state of the {\it t-J} model
is given by the Gutzwiller wave function \cite{KY}.

As in the previous study of thermodynamics \cite{KK95J,KK96,Pol,SS},
we use the Sutherland model with the  SU($K$,1) internal symmetry as an
auxiliary, and take the limit $\beta \rightarrow \infty$ of the coupling
parameter $\beta$. 
Then the particles crystallize with equal distance from each other,
and the resultant dynamics excluding phonons and uniform motion of the
center of gravity is mapped to that of the  SU($K$,1) {\it t-J} model
given by eq.(\ref{K-tj-hamiltonian}).
The merit of using the Sutherland model is that much more is known
about properties of the eigenfunctions than those for the {\it t-J}
model.
The Sutherland model with arbitrary internal symmetry is written in
the form
 %
\begin{equation}
{\cal H}
=
-\frac{1}{2 M}
\sum_{i=1}^N \frac{\partial^2}{\partial x_i^2} +
\frac{\pi^2}{L^2 M}
\sum_{i<j}
\frac{\beta \left( \beta - s_{ij}\right)}
     {\sin^2 \left[ \pi \left(x_i-x_j \right) /L \right] },
\label{CSmodel}
\end{equation}
where $L$ is the size of the system, and $s_{ij}$ represents
the exchange operator of coordinates of particles $i$ and $j$
\cite{Pol2}.
In order to reproduce the lattice model
we take the limit of large $\beta$ and $M$, keeping the ratio $t =
\beta/M$ fixed.
Here we take the lattice parameter $L/N$  as the unit of length.
Note that the symmetry of the wave function leads to the relation
$s_{ij} = -\tilde{P}_{ij}$
within the Hilbert space of identical particles.

Let us describe derivation of $N(Q,\omega)$ taking the SU(2,1) 
Sutherland model.
Generalization to the SU($K$,1) supersymmetry is straightforward.
The system has
$N_{{\rm h}}$ holes,
$N_{\uparrow}$  up-spin electrons
and $N_{\downarrow}$ down-spin ones,
whose coordinates are represented by
$x_i^{\rm h}$ for $i$-th hole,
$x_i^{\uparrow}$ for $i$-th up-spin electron
and $x_i^{\downarrow}$ for $i$-th down-spin electron.
We arrange them as $x \equiv (x_1,x_2,\cdots,x_N)=(x_1^{\rm
h},\cdots,x_{N_{{\rm h}}}^{\rm
h},x_1^{\uparrow},\cdots,x_{N_{\uparrow}}^{\uparrow},x_1^{\downarrow},\cdots
,x_{N_{\downarrow}}^{\downarrow}) \equiv (x^{{\rm
h}},x^{\uparrow},x^{\downarrow})$
and assume that $N_{{\rm h}}$ is even, and both $N_{\uparrow}$ and
$N_{\downarrow}$ are odd. Then the ground state is non-degenerate.
The interval $I= [1,N]$ denotes $ \{ i \in
\mbox{\bf Z} | 1 \le i \le N \}$ for a positive integer $N$. We
define $I_{\rm h} =[1,N_{{\rm h}}], I_{\uparrow}=[N_{{\rm
h}}+1,N_{{\rm
h}}+N_{\uparrow}] $ and $I_{\downarrow}=[N_{{\rm h}}+N_{\uparrow}+1,N]
 $.
Without loss of generality we assume $N_{\downarrow} \ge
N_{\uparrow}$.
The wave function of the ground state for a set of $N_{{\rm
h}},N_{\uparrow},N_{\downarrow}$ is
given by,
\begin{equation}
\Psi_{\rm GS}
=
\prod_{i<j \in I} \left( 1-\frac{z_j}{z_i} \right)^\beta
\prod_{\sigma=\uparrow,\downarrow}
\prod_{i<j \in I_{\sigma}} \left( 1-\frac{z_j}{z_i} \right),
\label{gs}
\end{equation}
where the complex coordinates $z=(z_1,\cdots,z_N)$ are related
to the
original ones $x=(x_1,\cdots,x_N)$ by
$z_j = \exp(2 \pi i x_j /L)$.

The spectrum of the Sutherland model is conveniently analyzed with use
of a similarity transformation generated by
${\cal O} =  \prod_{i<j \in I} \left( 1-z_j/z_i \right)^\beta \prod_{i =
\in
I} z_i^{(N_{\downarrow}-1)/2}$.
The transformed Hamiltonian
$ \hat{\cal H}  =  {\cal O}^{-1} {\cal H } {\cal O} -e_0 $
describes the excited states relative to the spinless particle model
with the ground-state energy $e_0 = \pi^2 \beta^2 N(N^2-1)/6 M L^2$.
We obtain
\begin{equation}
\hat{{\cal H}}
=
\frac{1}{2 M}
\left( \frac{2 \pi }{L} \right)^2
\sum_{i=1}^N
\left(
\hat{d}_i + \beta \frac{N-1}{2}-
\frac{N_{\downarrow}-1}{2}
\right)^2. 
\end{equation}
Here $\hat{d}_i$ is called the Cherednik-Dunkl operator\cite{Mac} and is
given by
\begin{eqnarray}
\hat{d}_i
= z_i \frac{\partial}{\partial z_i}
+ \beta \sum_{k<i}\frac{z_i}{z_i-z_k} \left(1-s_{ik} \right)
\nonumber \\
+ \beta \sum_{i<k} \frac{z_k}{z_i-z_k} \left(1-s_{ik}\right)
+ \beta(1-i).
\label{dunkl}
\end{eqnarray}
It is known that  $\hat{d}_i$ can be diagonalized simultaneously with
real eigenvalue $\bar{\lambda}_i$.
The resultant eigenfunction are polynomials $E_\lambda (z;\beta)$ of
$z_i$ such as $\prod_i z_i^{\lambda_i}+\ldots$, and are called
non-symmetric Jack polynomials\cite{Mac}.
Since we are dealing with identical particles,
eigenfunction should satisfy the following conditions of the SU(2,1)
supersymmetry:\\
(i) symmetric with respect to exchange between  $z_i^{\rm h}$'s; \\
(ii) anti-symmetric with respect to exchange between  $z_i^{\sigma}$'s
with the same $\sigma$.

By taking linear combination of $E_\lambda (z;\beta)$, we can
construct a polynomial $K_\lambda (z;\beta)$ with the SU(2,1)
supersymmetry.
The latter belongs to a family called the Jack polynomial with
prescribed symmetry \cite{BF,Dunkl98,YK}.
We specify the set of momenta as
$\left(\lambda^{\rm h},\lambda^{\uparrow},\lambda^{\downarrow} \right)
\equiv (\lambda^{\rm h}_1,\cdots, \lambda^{\rm h}_{N_{\rm
h}},\lambda^{\uparrow}_1,\cdots,\lambda^{\uparrow}_{N_{\uparrow}},
\lambda^{\downarrow}_1,\cdots,\lambda^{\downarrow}_{N_{\downarrow}})$
with
$\lambda^{\rm h}_1 \ge \cdots  \ge \lambda^{\rm h}_{N_{{\rm h}}},
\lambda^{\sigma}_1 > \cdots  > \lambda^{\sigma}_{N_{\sigma}}  \ (\sigma=\uparrow,\downarrow)$. 
In this way we can parameterize  $K_{\lambda} (z; \beta)$ by using
$\lambda \equiv \left(\lambda^{\rm
h},\lambda^{\uparrow},\lambda^{\downarrow} \right)$.
At the ground state we have $\lambda =\lambda_{\rm GS} = (\mu^{\rm
h},\mu^\uparrow,\mu^\downarrow)$
 with
$
\mu^{\rm h} = ((N_{\downarrow}-1)/2, (N_{\downarrow}-1)/2, \cdots,
(N_{\downarrow}-1)/2),\
\mu^\uparrow =
((N_{\uparrow}+N_{\downarrow})/2-1,(N_{\uparrow}+N_{\downarrow})/2-2,\cdots,
(N_{\downarrow}-N_{\uparrow})/2) $ and $
\mu^\downarrow = (N_{\downarrow}-1,N_{\downarrow}-2,\cdots, 0).
$

We define the inner product
for functions $f(z)$ and $g(z)$ in $N$-complex variable
$z=(z_1,\cdots,z_N)$
as follows:
\begin{equation}
\langle f,g \rangle_0
= \prod_{j=1}^N \oint_{ | z_j|=1}
\frac{d z_j}{2\pi i z_j}
\overline{f(z)} g(z)
\prod_{k<l \in I } | z_k-z_l |^{2 \beta},
\label{inner}
\end{equation}
where $\overline{f(z)}$ denotes the complex conjugation of $f(z)$.
As is clear from the definition of the transformation ${\cal O}$,
$K_\lambda (z;\beta)$ is orthogonal with respect to the above inner
product.
In order to derive the norm of $K_\lambda (z;\beta)$, we generalize
the procedure taken in refs.\cite{YK,Takemura,pr}.
The result of lengthy calculation is given by
\begin{equation}
\langle K_{\lambda}, K_{\lambda}\rangle_0
=
N_{{\rm h}} ! N_{\downarrow}! N_{\uparrow}!
\rho_{\lambda}^{-1}
\langle E_{\lambda}, E_{\lambda}\rangle_0,
\end{equation}
where $\langle E_{\lambda}, E_{\lambda}\rangle_0$ denotes the norm of 
the
non-symmetric Jack polynomials.
We refer to the literature \cite{Mac} for the explicit form of the norm 
since
it requires many lengthy combinatorial quantities.
The quantity $\rho_{\lambda} =\rho_{\lambda}^{\rm h}
\rho_{\lambda}^{\uparrow}
\rho_{\lambda}^{\downarrow}$ is given by
\begin{equation}
\rho_{\lambda}^{\rm h}
=
\prod_{i<j \in I_{\rm h} }
\frac{\bar{\lambda}_i-\bar{\lambda}_j+\beta}
     {\bar{\lambda}_i-\bar{\lambda}_j} ,
\ \
\rho_{\lambda}^{\sigma}
=
\prod_{i<j \in I_\sigma}
\frac{\bar{\lambda}_i-\bar{\lambda}_j-\beta}
     {\bar{\lambda}_i-\bar{\lambda}_j}.
\end{equation}

In the lattice model, the density correlation of particles is the same
as
that of holes
because of the completeness relation
$\sum_{\sigma=\uparrow,\downarrow} X_i^{\sigma\sigma}+X_i^{00}=1$.
In the Sutherland model, on the other hand, the correlation function
of the sum of hole-density and particle-density gives the phonon
spectrum. 
It can be shown that the intensity of the phonon correlation function is
smaller than the spin correlation
function by a factor of $O(\beta^{-1})$. 
In the leading order in $\beta^{-1}$, the correlation function of holes 
is the
same as that of particles. 
In view of this situation we consider
 the power sum $p_m = \sum_{i \in I_{\rm h}} z_i^m$ for holes.
The hole-density $n_Q$  can be expressed as
$ n_Q = p_m/ \sqrt{N}$ with $Q=2 \pi m/N$ \cite{Ha,Lesage}.
In order to calculate $N(Q,\omega)$, we need to know
the expansion coefficient $c_{\lambda}$ which appears in
\begin{equation}
p_m \Psi_{\rm GS}
=
\sum_{\lambda}
c_{\lambda} K_{\lambda} (z;\beta)
\prod_{i<j\in I} \left(1- \frac{z_j}{z_i} \right)^{\beta}.
\label{K-expansion}
\end{equation}
It is difficult to derive $c_\lambda$ for general values of $m$.
However, there occurs drastic simplification for the small momentum
region with
$m \leq (N_{\uparrow}-1)/2$.
In this region one can show that $p_m \Psi_{\rm GS}$ has only charge
excitations due to
the triangular structure of the polynomial $K_{\lambda}(z; \beta)$ 
\cite{KK95}.

Without spin excitations, we have the explicit
expression of $K_{\lambda}(z; \beta)$
following Baker and Forrester\cite{BF} as follows:
\begin{eqnarray}
K_{\lambda}(z; \beta)
&=&
J_{\nu}^{\alpha}(z_{\rm h})
\prod_{i \in I} z_i^{(N_\downarrow-1)/2}
\prod_{\sigma=\uparrow,\downarrow}
\prod_{k<l\in I_\sigma}
\left( 1 - \frac{z_l}{z_k} \right),
\label{BakerForr}
\end{eqnarray}
where $\alpha = 2 +1/\beta$ and
$J_{\nu}^{\alpha}(z_{\rm h})=J_{\nu}^{\alpha}(z_1,\cdots,z_{N_{{\rm =
h}}})$
is the
symmetric Jack polynomial \cite{Mac}.
The charge excitations are characterized by the set $\nu =
(\nu_1,\cdots,\nu_{N_{\rm h}})$, which is
related to $\lambda$ by
$\lambda = (\nu,0,0)+\lambda_{\rm GS}$.

Since $p_m$ can be expanded by the symmetric Jack polynomials
\cite{Hanlon},
we can determine the coefficient $c_{\lambda}$ in
eq.(\ref{K-expansion}).
With use of a square $s =(i,j)$ in the Young diagram $\nu$, we
introduce the notations
$$
[ 0' ]^{\alpha}_{\nu}
=
\prod_{ s \in \nu \atop{s \ne (1,1)}}
[\alpha  a'_{\nu}(s) -l'_{\nu}(s) ],\
h^{\alpha}_{\nu}
=
\prod_{s \in \nu} [\alpha (a_{\nu}(s) +1) + l_{\nu}(s) ] ,
$$       
where $a_\nu (s)=\nu_i - j$ is the arm-length, $a'_\nu (s)= j-1$ is
the arm-colength, 
$l_{\nu}(s)=\nu'_j-i$ with $\nu'_j$ the length of the column $j$
 is the leg-length,
and $l'_{\nu}(s)=i-1$ is the leg-colength.
We then obtain the dynamical structure factor of the Sutherland
model with the coupling parameter $\beta$ as follows:
\begin{eqnarray}
N \left( Q,\omega ;\beta \right)
& = &
\frac{\alpha^2  m^2}{N}
\sum_{ \nu } \delta_{m,|\nu|}
\delta(\omega-E_{\lambda}+E_{\lambda_{{\rm GS}}})
\nonumber \\
& & \times
\frac{ \langle K_{\lambda},K_{\lambda}\rangle_0}
     {\langle K_{\lambda_{\rm GS}},K_{\lambda_{\rm GS}}\rangle_0}
\left(
\frac{ [ 0']^{\alpha}_{\nu} }{h^{\alpha}_{\nu}}
\right)^2,
\label{finitenqw}
\end{eqnarray}
where $Q= 2 \pi m/N$ and
$
|\nu| = \sum_{j=1} ^{N_{\rm h}} \nu_j .
$
The excitation energy is given by
\begin{equation}
E_{\lambda}-E_{\lambda_{{\rm GS}}}
=
\frac{\beta}{2M} \sum_{i=1}^{N_{\rm h}}
[ \alpha \nu_i^2 + (N_{\rm h} +1-2 i)\nu_i ] .
\end{equation}

By taking the limit $\beta\rightarrow\infty$ with the ratio $t =
\beta/M$ kept fixed,
we can derive $N(Q,\omega)$ of the $t$-$J$ model
$(\ref{tj-hamiltonian})$.
In this limit $N(Q,\omega)$ simplifies dramatically by the following
fact:
The coefficient $[ 0']^{\alpha}_{\nu}$ in eq.$(\ref{finitenqw})$
vanishes if the partition $\nu$ contains $s=(3,2)$.
Then $N(Q,\omega)$ is determined by three parameters which are
related directly with momentum of the
elementary excitations: two holons and one antiholon.
We emphasize that $N(Q,\omega)$ does not depend on the spin
polarization as long as $Q$ is smaller than the smallest Fermi wave 
number
of all
electrons.

Derivation of dynamical charge structure factor for SU($K$,1) model
proceeds in a similar manner.
We write the momenta of $K$ holons and one antiholon
by $\lambda_1,\cdots,\lambda_K$ and $\lambda'$, respectively.
Then $N(Q,\omega)$ for finite $N$ and arbitrary $K$ can be expressed as
follows:
\begin{eqnarray}
\lefteqn{
N(Q,\omega)
=
B_0 m^2 \sum_{\lambda_1 \geq \cdots \geq\lambda_K,\lambda'}
\delta_{m, ||\lambda||}
\delta\left( \omega - E_{\lambda \lambda'} \right)
}
\nonumber \\
&  &
\times
\frac{\Gamma( \lambda' )\Gamma(N_{\rm h} - \lambda' + K) }
     {\Gamma( \lambda' +1-K )\Gamma( N_{\rm h} - \lambda' +1) }
\nonumber \\
&  &
\times
\prod_{j=1}^K
\frac{1}
{\left(K \lambda_j+ \lambda' -j \right)
\left( K(\lambda_j-1) + \lambda' -(j-1)  \right)}
\nonumber \\
& &
\times
\prod_{j=1}^K 
\frac{\Gamma(\lambda_j -(j-1)/K )\Gamma( \lambda_j +(N_{\rm h} -j +1 )/K
) }
{\Gamma(\lambda_j -(j-K)/K ) \Gamma( \lambda_j + (N_{\rm h} -j +K )/K ) 
}
\nonumber\\
& &
\times
\prod_{1 \leq i < j\leq K}
\left[
\frac{\Gamma(\lambda_i - \lambda_j + 1 - (i-j)/K)}
     {\Gamma(\lambda_i - \lambda_j + 1 - (i-j+1)/K )}
\right.
\nonumber \\
& &
\times
\left.
\frac{\Gamma(\lambda_i - \lambda_j - (i-j-1)/K )}
     {\Gamma(\lambda_i - \lambda_j - (i-j)/K)}
\right], 
\label{finiteversion}
\end{eqnarray}
where $||\lambda|| = \lambda' +\sum_{j=1}^K (\lambda_j -1)$ and $B_0 =2 \pi A_0 K^{1-K}K!/N$.
The excitation energy is given by
\begin{eqnarray}
E_{\lambda \lambda'}
&=&\frac{t}{2}\left(\frac{2 \pi}{N}\right)^2
\{ \lambda'(N_{\rm h} +K-\lambda')
\nonumber \\
&+& \sum_{j=1}^K [ (\lambda_j-1) ( N_{\rm h} + K(\lambda_j+1)-(2j-1) ) 
] \}.
\end{eqnarray}
In the special case of $\lambda_K = 0$ where less than $K$ holons are
excited, we need to modify
the form of eq.(\ref{finiteversion}).
However, this case can be neglected in the thermodynamic limit.
We have checked the validity of eq.(\ref{finiteversion}) together with
the
modified one
in the case of $K=2$ by comparison with numerical results up to 
$N=16$.  

Following the same procedure as that in the spinless Sutherland
model\cite{Ha,Lesage},
we obtain the expression (\ref{kfn})  in the thermodynamic limit.
The present derivation makes it clear that the spin-charge separation
persists in systems with finite size. 
In the SU($K$,1) model, $\alpha$ in eq.(\ref{BakerForr})
is given by $\alpha =K+\beta^{-1}$, which  
tends to $K$  in the limit of $\beta\rightarrow\infty$.
Since $1/\alpha$ can be regarded as the coupling constant in
the charge sector of the fractional statistics,
$N(Q,\omega)$ given by eq.(\ref{kfn}) has the
apparent coupling constant $1/K$.

The authors would like to thank Y. Kato for valuable discussions.
T.Y. and Y.S. wish to acknowledge the support of the CREST from the 
Japan
Science and Technology Corporation.
T.Y. wishes to thank the support of the Visitor Program of the MPI-PKS.

\end{multicols}
\end{document}